\documentclass[twocolumn]{revtex4-1}

\usepackage{latexsym}
\usepackage{amssymb}
\usepackage{amsmath}
\usepackage{amsthm}
\usepackage{amsfonts}
\usepackage{ifthen}
\usepackage{revsymb}
\usepackage{yfonts}
\usepackage{graphicx}
\usepackage{algpseudocode}
\usepackage{bbm}
\usepackage{multirow}
\usepackage{gensymb}
\usepackage{epstopdf}
\usepackage{listings}
\usepackage{color}
\definecolor{mygreen}{rgb}{0,0.6,0}
\definecolor{mygray}{rgb}{0.5,0.5,0.5}
\definecolor{mymauve}{rgb}{0.58,0,0.82}

\usepackage{afterpage}
\usepackage{verbatim}
\usepackage{soul}
\usepackage{lineno, blindtext}
\usepackage{footnote}
\usepackage{afterpage}
\usepackage{algpseudocode}
\usepackage[normalem]{ulem}
\usepackage{algorithm}
\usepackage{algpseudocode}
\usepackage{mathtools}
\DeclarePairedDelimiter{\ceil}{\lceil}{\rceil}
\usepackage[colorlinks = true]{hyperref}
\usepackage{longtable}
\usepackage{xcolor}
\definecolor{darkred}  {rgb}{0.5,0,0}
\definecolor{darkblue} {rgb}{0,0,0.5}
\definecolor{darkgreen}{rgb}{0,0.5,0}
\usepackage{amsmath}
\newcommand{\tens}[1]{%
  \mathbin{\mathop{\otimes}\limits_{#1}}%
}
\hypersetup{
  urlcolor   = blue,         % color of external links
  linkcolor  = darkblue,     % color of internal links
  citecolor  = darkgreen,    % color of links to bibliography
  filecolor  = darkred       % color of file links
}

\newcommand{\be}{\begin{equation}}
\newcommand{\ee}{\end{equation}}
\newcommand{\bq}{\begin{eqnarray}}
\newcommand{\eq}{\end{eqnarray}}
\newcommand{\bea}{\begin{eqnarray}}
\newcommand{\eea}{\end{eqnarray}}
\newcommand{\ba}{\begin{align}}
\newcommand{\ea}{\end{align}}

\newcommand{\ket}[1]{ | \, #1 \rangle}
\newcommand{\bra}[1]{ \langle #1 \,  |}

\definecolor{mygray}{gray}{0.6}

\newcommand{\beginsupplement}{%
        \setcounter{table}{0}
        \renewcommand{\thetable}{S\arabic{table}}%
        \setcounter{figure}{0}
        \renewcommand{\thefigure}{S\arabic{figure}}%
     }
\definecolor{mygray}{gray}{0.9}

\begin{document}
\title{Efficient quantum algorithm for solving travelling salesman problem: An IBM quantum experience}

\author{Karthik Srinivasan$^1$}
\thanks{These authors contributed equally to this work.}
\author{Saipriya Satyajit$^2$}
\thanks{These authors contributed equally to this work.}
\author{Bikash K. Behera$^3$}
\author{Prasanta K. Panigrahi$^3$}
\affiliation{$^1$Department of Physics, Indian Institute of Technology Madras, Chennai 600036, India}
\affiliation{$^2$Department of Physics, Indian Institute of Technology Bombay, Mumbai 400076, India}
\affiliation{$^3$Department of Physical Sciences, Indian Institute of Science Education and Research Kolkata, Mohanpur 741246, West Bengal, India}

\maketitle
\noindent
 
\textbf{The famous Travelling Salesman Problem (TSP) is an important category of optimization problems \cite{tsp_mottnat2017} that is mostly encountered in various areas of science and engineering. Studying optimization problems motivates to develop advanced techniques more suited to contemporary practical problems \cite{tsp_WesterlundSD1998,tsp_DebMOOP2009,tsp_Wangarxiv2014,tsp_Fiorettoarxiv2016,tsp_ZhaoJOTA2018,tsp_ButenkoJOTA2003}. Among those, especially the NP hard problems provide an apt platform to demonstrate supremacy of quantum over classical technologies in terms of resources and time. TSP is one such NP hard problem in combinatorial optimization \cite{tsp_Daiarxiv2018,tsp_HoffmanJCAM2000} which takes exponential time order for solving by brute force method. Here we propose a quantum algorithm to solve the travelling salesman problem using phase estimation technique. We approach the problem by encoding the given distances between the cities as phases. We construct unitary operators whose eigenvectors are the computational basis states and eigenvalues are various combinations of these phases. Then we apply phase estimation algorithm to certain eigenstates which give us all the total distances possible for all the routes. After obtaining the distances we can search through this information using the quantum search algorithm for finding the minimum \cite{tsp_Durrarxiv1996} to find the least possible distance as well the route taken. This provides us a quadratic speedup over the classical brute force method for a large number of cities. In this paper, we illustrate an example of the travelling salesman problem by taking four cities and present the results by simulating the codes in the IBM's quantum simulator.}

Travelling Salesman Problem (TSP) is a classic optimization problem \cite{tsp_mottnat2017} in the field of computer science. It belongs to an intriguing class of `hard' optimization problems called NP hard \cite{tsp_Kumlander2008,tsp_Herrnat2017}. The problem involves a salesman who has to travel N cities, visiting each city once and reaching ultimately at the same city where he started. This cycle of visiting each city once, where each city represents a unique vertex in a graph, and returning to the starting city is known as a Hamiltonian cycle \cite{tsp_berge1973graphs}. Each city is connected to other cities with a specific cost associated to each connection. The cost gives an idea of how difficult it is to take the corresponding route. The aim of the salesman is to minimize the cost of travel, satisfying the above described conditions. Even if we break the travelling salesman problem into smaller components, each component will be at least as complex as the initial problem. This is why it belongs to the class of NP hard problems. The most expensive and simplest classical solution to the problem is to find the solution by brute force method. However, the problem becomes impossible to solve when a large number of cities are taken. For N cities, (N-1)! possible iterations are needed to search for the solution, which shoots up very fast as N increases. Other classical approaches to solve the problem include branch and bound algorithms \cite{tsp_LawlerOR1966,tsp_PadbergORL1987,tsp_LittleOR1963,tsp_HeldOR1970}, heuristics \cite{tsp_LinOR1973,tsp_GeemSIM2001,tsp_LaporteDAM1990,tsp_KargMS1964} and other methods \cite{tsp_BhideIEEE1993,tsp_ShaelaieASC2014,tsp_ShwetaIJST2017}. Using branch and bound algorithms the problem has been solved for around 86,000 cities, but the success in branch and bound algorithms depends on certain factors which if not satisfied give us the same complexity as the brute force method. Heuristics approach is based on providing a set of rules on optimal selection of next city to travel. But this does not give optimal solution in every case as heuristics result in approximations.

With the advent of the era of quantum technologies possibilities of solving this problem with quantum computers has come to the limelight with the aim to tackle a much bigger problem of proving P = NP class. Several quantum algorithms \cite{tsp_Keiuarxiv2018} have also been proposed aiming at the same. Goswami \emph{et al.} \cite{tsp_Goswamiarxiv2004} have presented a framework for efficiently solving the approximate travelling salesman problem. A quantum heuristic algorithm has been proposed by Bang et al. \cite{tsp_BangJKPS2012} to solve the traveling salesman problem by generalizing the Grover search. Moylett \emph{et al.} \cite{tsp_MoylettPRA2017} have given a proof of the quantum quadratic speed up for the Travelling Salesman Problem for bounded-degree graphs. The above mentioned algorithms work only when certain conditions are satisfied, however our algorithm combined with the quantum algorithm for finding the minimum by Durr and Hoyer \cite{tsp_Durrarxiv1996} gives a quadratic speedup over the classical brute force method without further conditions on the problem. 

The classical algorithms to solve the problem take input in the form of a matrix say A such that $[A]_{ij} =  \phi_{ij}$, where $\phi_{ij}$ is the cost/distance/time or any other quantity taken to travel from city i to city j. This quantity for the overall journey has to be minimized. Without loss of generality, we take the quantity as cost in the current work. In the problem, the main motivation to take input as phases stems from the following two facts; first the matrix made of the given distances using the above procedure is not unitary in general which implies that the implementation and manipulation of the operator is not possible on a quantum computer. Second, the phases will get added when we multiply them or take tensor products of states with these phases as coefficients, that is the distances will get added as phases which is required for the search. Hence we represent the input as a martix B, where $B_{ij} = e^{i(\phi_{ij})}$.

\begin{figure}[h!]
    \centering
    \includegraphics[scale = 1.75]{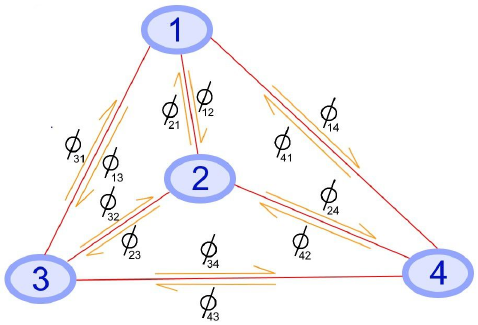}
    \caption{The figure illustrates the Travelling Salesman Problem for four cities. The cities are represented by circles named 1, 2, 3 and 4 and the cost for each path connecting the cities are labelled on them. Here, $\phi_{ij}$ represents the cost of travel from city $i$ to city $j$. This is the most general case of travelling salesman where the cost for travelling from city $i$ to city $j$ is not the same as the cost for travelling from city $j$ to city $i$.}
\label{tsp_Fig1}
\end{figure}

Next major step in our algorithm is phase estimation which is the backbone of our algorithm. In fact the solution of Travelling Salesman Problem (TSP) we present here seems to be a perfect application of phase estimation algorithm. The next step in the algorithm is to construct unitaries $U_j$ from the matrix B as
\begin{equation}
\label{tsp_eq1}
[U_j]_{kk} =\frac{1}{\sqrt{N}}[B]_{jk}, 
\end{equation}
where N is the number of cities and j, k $\geq$ 0 and j, k $\in [1,N]$. The rest of the elements in the matrix $U_j$ are initialized to zero. It is cleary observed that $U_j$ is a diagonal unitary matrix. Using these unitaries, we now create another unitary matrix which is the tensor products of all the unitaries; $U = U_1\tens{}U_2\tens{}...U_N$. The eigenvalues of this unitary matrix U, are estimated using the phase estimation algorithm. The phases can be easily normalized to be bound within 0 and $2\pi$ once we know the range of distances between the cities which is given to us in the problem. Here, U is a diagonal matrix since it is a tensor product of N diagonal matrices. This means that the eigenstates of this matrix U are computational basis states with eigen values as the corresponding diagonal elements. Hence there will be (N-1)! diagonal elements of interest to us out of the $N^N$ elements. These elements will have the total cost of the (N-1)! possible Hamiltonian cycles as phases. This implies that there are (N-1)! eigenstates of U with eigenvalues being the total cost of the corresponding Hamiltonian cycle as phase. It is easy to calculate the position of these elements in this matrix U which will have this ``useful" information. Since for a given number of cities, we know the location of the elements with total cost, we can prepare computational basis eigenstates corresponding to the location of each of these elements and extract the phase using phase estimation. We get the phases in form of binary output from phase estimation algorithm, then we can easily perform the quantum algorithm for finding the minimum \cite{tsp_Durrarxiv1996} to find the minimum cost and the corresponding route that is to be taken for that particular cost. 

\begin{figure}[h!]
    \centering
    \includegraphics[scale = 0.5]{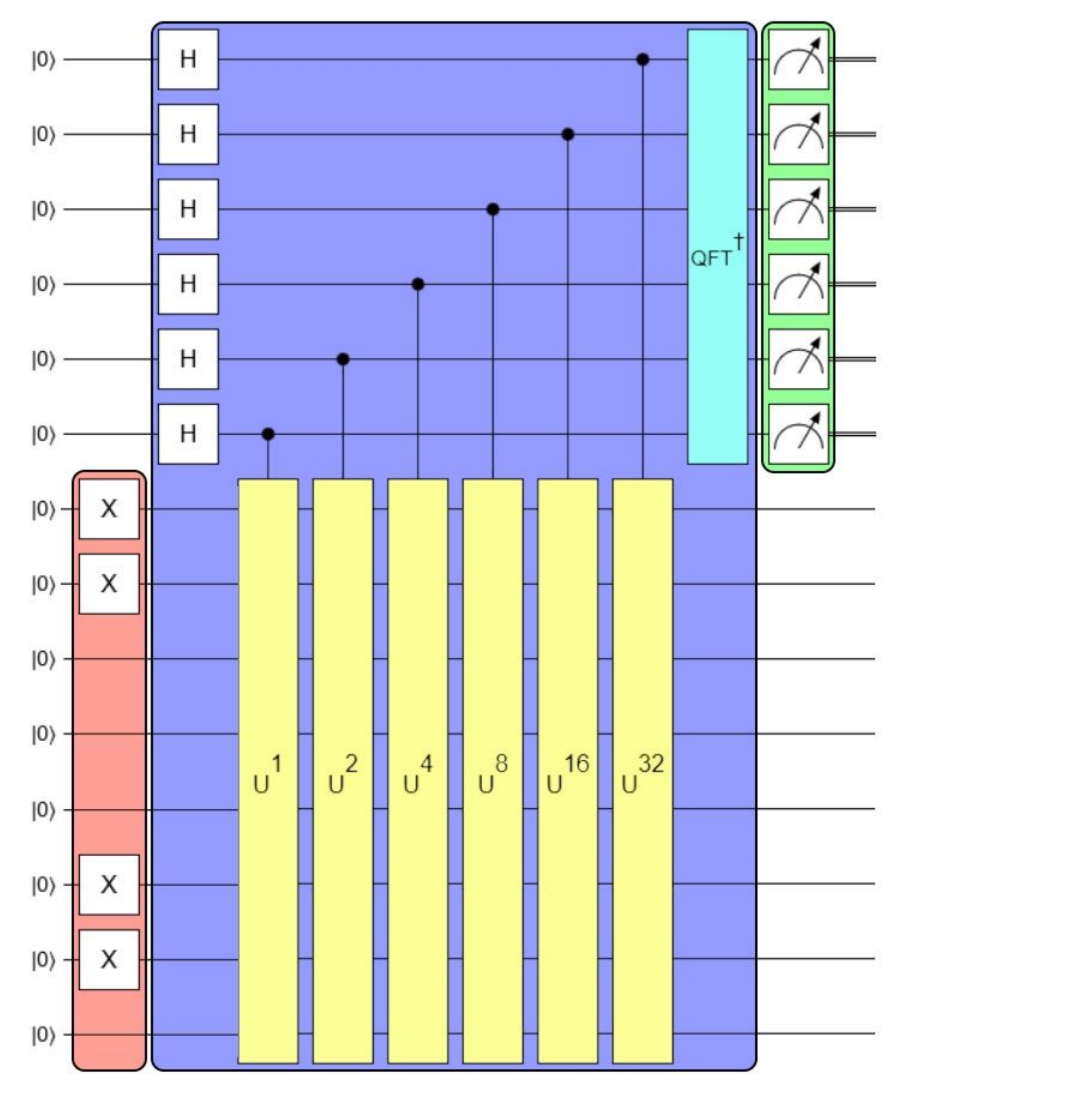}
    \caption{The above figure shows the quantum circuit for phase estimation of eigenstate $\ket{11000110}$ which corresponds to the route (Hamiltonian cycle) $1\rightarrow2\rightarrow3\rightarrow4\rightarrow1$. We have taken here six  qubits for estimating the phase and rest eight qubits for preparing the corresponding eigenstate (all of which are initialized in the $\ket{0}$ state) by the use of pauli X gates on specific qubits. The unitary U is the one described earlier i.e. the tensor product unitaries $U_1\tens{}U_2\tens{}...U_N$. The part of the circuit in red represents initialization of the eigenstates. The part of the circuit in blue performs the phase estimation method.}
    \label{tsp_Fig2}
\end{figure}

In the travelling salesman problem with four cities shown in the Fig. \ref{tsp_Fig1}, we can take a total of six routes which satisfy the conditions for being a Hamiltonian cycle. If the cost of travelling from city $i$ to city $j$ is the same as that of city $j$ to city $i$ (the symmetric case) then we will get only three routes with unique total costs. We have taken the symmetric case for implementing on IBM quantum experience. We have implemented this problem using the circuit given in the Fig. \ref{tsp_Fig2} on the IBM quantum experience custom topology. However in the simulation we took six qubits for phase estimation instead of four apart from the eigenstate qubits. We repeated this circuit six times, once for each eigenstate. The eigenstates corresponding to the different routes, the theoretical expectation for the values of $\phi_{ij}$ and the experimental observations are given in the table \ref{tsp_table1}.

\begin{table}[H]
\centering
\caption{The result of simulation for (4-1)! eigenstates and the theoretical expectations are presented in the T.bael The values taken as costs for travelling are given in the Supplementary Information Section}
 \begin{tabular}{c c c c} 
 \hline
 \hline
 SL No. & Eigenstate & Theoretical & Experimental \\ [0.5ex] 
 \hline\hline
 1. & 11000110 & 100100 & 100000 \\ 
 2. & 01101100 & 100100 & 101000 \\
 3. & 10001101 & 100000 & 010000 \\
 4. & 01110010 & 100000 & 010000 \\
 5. & 11100001 & 011000 & 101000 \\ 
 6. & 10110100 & 011000 & 010000 \\ [1ex] 
 \hline
 \hline
 \end{tabular}
 \label{tsp_table1}
\end{table}

Even though the experimental results do not exactly coincide with the theoretical expectations, we can rectify this by taking more qubits for phase estimation, which is its inherent property. To successfully obtain the phase accurate up to n bits with probability of success at least $1-\epsilon$, the number of qubits we need for phase estimation (t) is given by the following expression,

\begin{equation}
    t =  n + \ceil*{log\Bigg(2+\frac{1}{2\epsilon}\Bigg)}
\end{equation}

%*********************CONCLUSION************************************

Using the proposed algorithm, we are able to create a database of all possible routes that can be taken along with the distance of each. If one devices a quantum algorithm to find the minimum element in an unsorted array, which is faster than the one we currently have, then we can use that algorithm to find the minimum. This gives our algorithm a flexibility, which can be exploited in the future to solve the travelling salesman problem much more efficiently.

%*****************************************************************
%*********************Dealing with an exception*******************
%*****************************************************************

Even though our algorithm deals with a very general case, there are certain cases which cannot be directly solved using our algorithm. These are the cases where there are restrictions on routes connecting cities. For instance, city i does not have a route connecting it to city j. This can be thought of as the distance between those cities being infinite. Since our algorithm requires distances that can be normalized such that the total distance for the longest route is less than $2\pi$, this does not bode well for us. Fortunately, there is a way to deal with this exception. We can take the distance between the concerned cities as a very large distance such that the routes containing the path connecting city i and j will have a total distance which will certainly exceed the minimum distance. 

{\bf Supplementary Information} is available in the online version of the paper.

{\bf Acknowledgments}
B.K.B. is financially supported by DST Inspire Fellowship. S.S. and K.S. acknowledge the support of HBCSE and TIFR for conducting National Initiative on Undergraduate Sciences (NIUS) Physics camp. We are extremely grateful to IBM team and IBM QE project. The discussions and opinions developed in this paper are only those of the authors and do not reflect the opinions of IBM or IBM QE team.

{\bf Author contributions}
K.S. and S.S. contributed equally to this work. K.S. and S.S. came up with the efficient algorithm to solve the TSP problem. K.S. and S.S. designed the quantum circuit and simulated using IBM quantum experience. K.S., S.S. and B.K.B. contributed to the composition of the manuscript. K.S., S.S. and B.K.B. has done the work under the guidance of P.K.P.  

{\bf Author information} The authors declare no competing financial interests. Correspondence and requests for materials should be addressed to P.K.P. (pprasanta@iiserkol.ac.in).

\pagebreak

\onecolumngrid

\section*{ Supplementary Information: Efficient Quantum Algorithm for solving Travelling Salesman problem}
\beginsupplement

\section{4 City example explanation}
The distance matrix  for the Travelling salesman problem consisting of four cities is given by

\begin{equation}
\label{tsp_eq3}
D = \frac{1}{2}
\left[{\begin{array}{cccc}
                   1&e^{i\phi_{12}}&e^{i\phi_{13}}&e^{i\phi_{14}}\\
                   e^{i\phi_{21}}&1&e^{i\phi_{23}}&e^{i\phi_{24}}\\
                   e^{i\phi_{31}}&e^{i\phi_{32}}&1&e^{\phi_{34}}\\
                   e^{i\phi_{41}}&e^{i\phi_{42}}&e^{i\phi_{43}}&1\\
\end{array} } \right] 
\end{equation}

where
$\phi_{12} = \phi_{21} = \pi/2,$
$\phi_{13} = \phi_{31} = \pi/8,$
$\phi_{14} = \phi_{41} = \pi/4,$
$\phi_{23} = \phi_{32} = \pi/4,$
$\phi_{24} = \phi_{42} = \pi/4,$
$\phi_{34} = \phi_{43} = \pi/8,$

and the unitaries $U_j$ with j = 1, 2, 3 and 4 are as follows
\begin{equation}
\label{tsp_eq4}
    U_1 = \ket{00}\bra{00} + e^{i\phi_{21}}\ket{01}\bra{01} + e^{i\phi_{31}}\ket{10}\bra{10} + e^{i\phi_{41}}\ket{11}\bra{11}
\end{equation}
\begin{equation}
\label{tsp_eq5}
    U_2 = e^{i\phi_{12}}\ket{00}\bra{00} + \ket{01}\bra{01} + e^{i\phi_{32}}\ket{10}\bra{10} + e^{i\phi_{42}}\ket{11}\bra{11}
\end{equation}
\begin{equation}
\label{tsp_eq6}
    U_3 = e^{i\phi_{13}}\ket{00}\bra{00} + e^{i\phi_{23}}\ket{01}\bra{01} + \ket{10}\bra{10} + e^{i\phi_{43}}\ket{11}\bra{11}
\end{equation}
\begin{equation}
\label{tsp_eq7}
    U_4 = e^{i\phi_{14}}\ket{00}\bra{00} + e^{i\phi_{24}}\ket{01}\bra{01} + e^{\phi_{34}}\ket{10}\bra{10} + \ket{11}\bra{11}
\end{equation}

We constructed the unitaries $U_j$ by decomposing it as follows;

\begin{equation}
\label{tsp_eq8}
U_j =
\left[{\begin{array}{cccc}
                   e^{ia}&0&0&0\\
                   0&e^{ib}&0&0\\
                   0&0&e^{ic}&0\\
                   0&0&0&e^{id}\\
\end{array} } \right] 
= 
\left[{\begin{array}{cc}
                   1&0\\
                   0&e^{i(c-a)}\\
\end{array} } \right]
\tens{}
\left[{\begin{array}{cc}
                   e^{ia}&0\\
                   0&e^{ib}\\
\end{array} } \right]
\left[{\begin{array}{cccc}
                   1&0&0&0\\
                   0&1&0&0\\
                   0&0&1&0\\
                   0&0&0&e^{i(d+c-a-b)}\\
\end{array} } \right] 
\end{equation}

Note that this is the unitary $U_j$, by putting specific values of a, b, c, d we can find decomposition for each $U_1,U_2,U_3,U_4$. For phase estimation we need controlled-($U_1\tens{}U_2\tens{}U_3\tens{}U_4$) which is same as ($C-U_1\tens{}C-U_2\tens{}C-U_3\tens{}C-U_4$), where $C-U_1$ represents controlled-$U_1$. For realizing controlled $U_j$ we implemented controlled each element in the decomposition of $U_j$ in equation \ref{tsp_eq8}.

\begin{figure}[H]
    \centering
    \includegraphics[scale = 0.4]{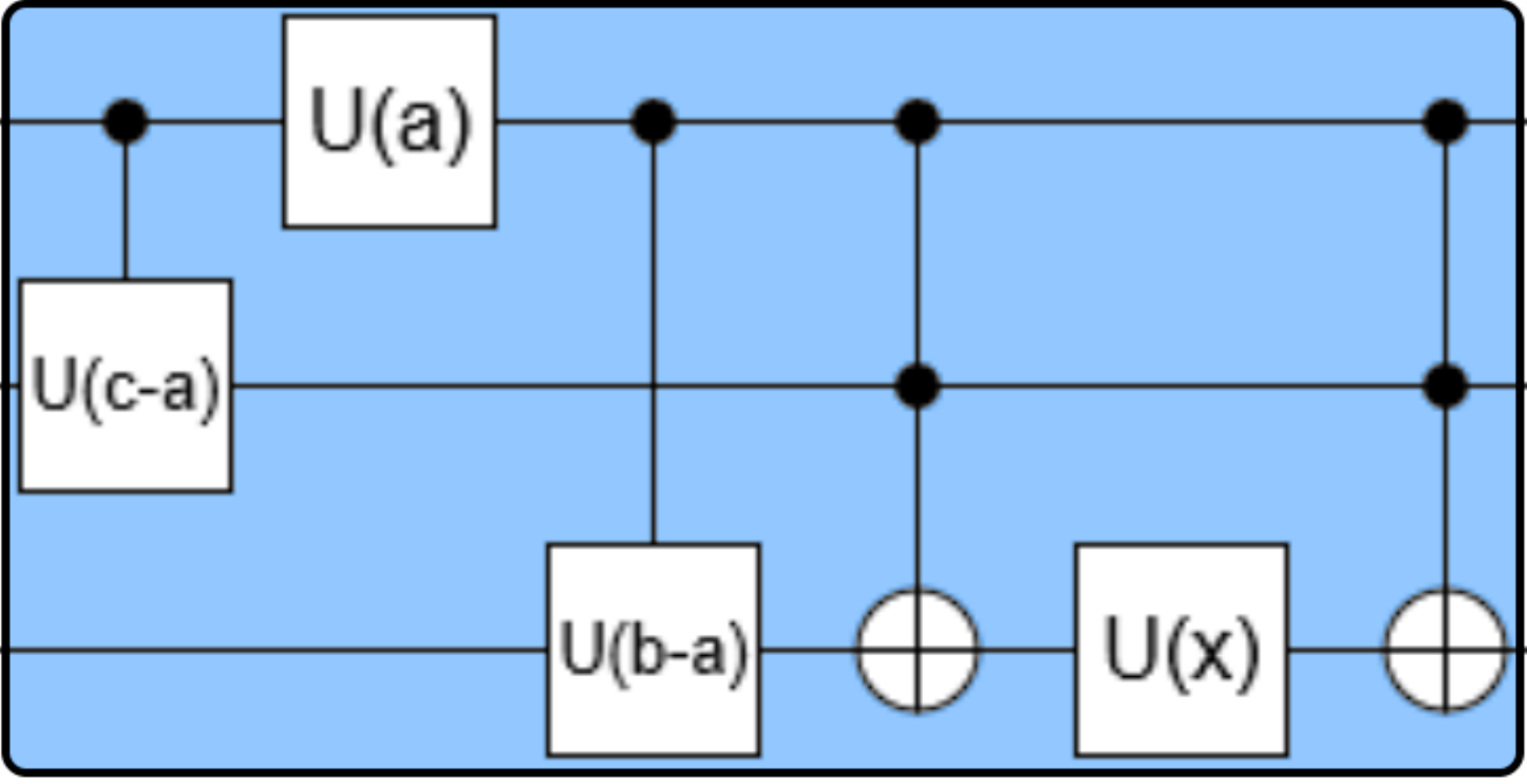}
    \caption{The figure illustrates our implementation of controlled $U_j$ gate with the use of single qubit unitaries U(a) as described earlier. Here we implement a controlled unitary (C-A) which is equivalent to implementing controlled unitaries that belong to the decomposition of A. In this figure, x = (d+c-a-b)/2}
    \label{tsp_Fig3}
\end{figure}

\section{The eigenstates} 

Fig. \ref{tsp_Fig2} shows the part of the circuit where we estimate the phase corresponding to the route going through cities 1, 2, 3 and 4 in the same order and returning to one. Starting from any other city but following the same order as mentioned will also give us the same eigenstate. The eigenstate for any route can be calculated as follows. In a particular route, if we are going from city $i$ to $j$, then each $i$ is uniquely mapped to $j$. Hence we can write $i$ as a function of $j$ i.e, $i(j)$. The eigenstate corresponding to that particular route is,
\begin{equation}
    \ket{\psi} = \tens{j} \ket{i(j) -1}
\end{equation}
where j goes from 1 to n.

Once these eigenstates are calculated, circuits similar to Fig. \ref{tsp_Fig1}, with eigenstate qubits initiated to the rest of the eigenstates, can be run in parallel. Then we can search through this database using the quantum search algorithm in the order of $O(\sqrt{(N-1)!)}$ steps to find the route with the least cost. If the cost of travelling from city i to city j is the same as city j to city i, then we can reduce the number of eigenstates by half of the original value. This means we will be able to search through the data in $O(\sqrt{((N-1)!)/2)}$ steps.

\section{Subroutines in IBM quantum Experience - custom topology}

In the simulation, we took advantage of the ``Add subroutine" under advanced option present in the custom topology \cite{tsp_sup_IBM1} and built these unitaries. The qasm code written for the simulation of the circuit (Fig. \ref{tsp_Fig4}) is given below.

\begin{lstlisting}[language=Python]
include "qelib1.inc";
qreg q[14];
creg c[6];

#Controlled unitary gate to implement C-U1, C-U2, C-U3 & C-U4
gate uni (a, b, c, d) x,y,z {
cu1 (c-a) x,y;
u1 (a) x;
cu1 (b-a) x,z;
ccx x,y,z;
cu1 ((d-c+a-b)/2) x,z;
ccx x,y,z;
cu1 ((d-c+a-b)/2) x,y;
cu1 ((d-c+a-b)/2) x,z;
}


#Controlled U = tensor product of U1, U2, U3 & U4
gate bigU (a, b, c, d, e, f, g, h, i, j, k, l) m,n,o,p,q,r,s,t,u {
uni (0,a,b,c) m,n,o;
uni (d,0,e,f) m,p,q;
uni (g,h,0,i) m,r,s;
uni (j,k,l,0) m,t,u;
}

#C-U with the values given
gate finU m,n,o,p,q,r,s,t,u {
bigU (pi/2,pi/8,pi/4,pi/2,pi/4,pi/4,pi/8,pi/4,pi/8,pi/4,pi/4,pi/8) m,n,o,p,q,r,s,t,u;
}

#C-U^2
gate finU2 m,n,o,p,q,r,s,t,u {
finU m,n,o,p,q,r,s,t,u;
finU m,n,o,p,q,r,s,t,u;
}

#C-U^4
gate finU4 m,n,o,p,q,r,s,t,u {
finU2 m,n,o,p,q,r,s,t,u;
finU2 m,n,o,p,q,r,s,t,u;
}

#C-U^8
gate finU8 m,n,o,p,q,r,s,t,u {
finU4 m,n,o,p,q,r,s,t,u;
finU4 m,n,o,p,q,r,s,t,u;
}

#C-U^16
gate finU16 m,n,o,p,q,r,s,t,u {
finU8 m,n,o,p,q,r,s,t,u;
finU8 m,n,o,p,q,r,s,t,u;
}

#C-U^32
gate finU32 m,n,o,p,q,r,s,t,u {
finU16 m,n,o,p,q,r,s,t,u;
finU16 m,n,o,p,q,r,s,t,u;
}

#All the controlled unitaries together
gate fU a,b,c,d,e,f,n,o,p,q,r,s,t,u {
finU f,n,o,p,q,r,s,t,u;
finU2 e,n,o,p,q,r,s,t,u;
finU4 d,n,o,p,q,r,s,t,u;
finU8 c,n,o,p,q,r,s,t,u;
finU16 b,n,o,p,q,r,s,t,u;
finU32 a,n,o,p,q,r,s,t,u;
}

#The rest are the gates for Inverse fourier transform
gate r2 x,y {
cu1 (-pi/2) x,y;
}
gate r3 x,y,z {
cu1 (-pi/4) x,z;
r2 y,z;
}
gate r4 w,x,y,z {
cu1 (-pi/8) w,z;
r3 x,y,z;
}
gate r5 v,w,x,y,z {
cu1 (-pi/16) v,z;
r4 w,x,y,z;
}
gate r6 u,v,w,x,y,z {
cu1 (-pi/32) u,z;
r5 v,w,x,y,z;
}


#The circuit
h q[0];
h q[1];
h q[2];
h q[3];
h q[4];
h q[5];
x q[6];
x q[8];
x q[9];
x q[11];
fU q[0],q[1],q[2],q[3],q[4],q[5],q[6],q[7],q[8],q[9],q[10],q[11],q[12],q[13];
h q[0];
r2 q[0],q[1];
h q[1];
r3 q[0],q[1],q[2];
h q[2];
r4 q[0],q[1],q[2],q[3];
h q[3];
r5 q[0],q[1],q[2],q[3],q[4];
h q[4];
r6 q[0],q[1],q[2],q[3],q[4],q[5];
h q[5];
measure q[0] -> c[0];
measure q[1] -> c[1];
measure q[2] -> c[2];
measure q[3] -> c[3];
measure q[4] -> c[4];
measure q[5] -> c[5];

\end{lstlisting}

\begin{figure}[H]
    \centering
    \includegraphics[scale = 0.6]{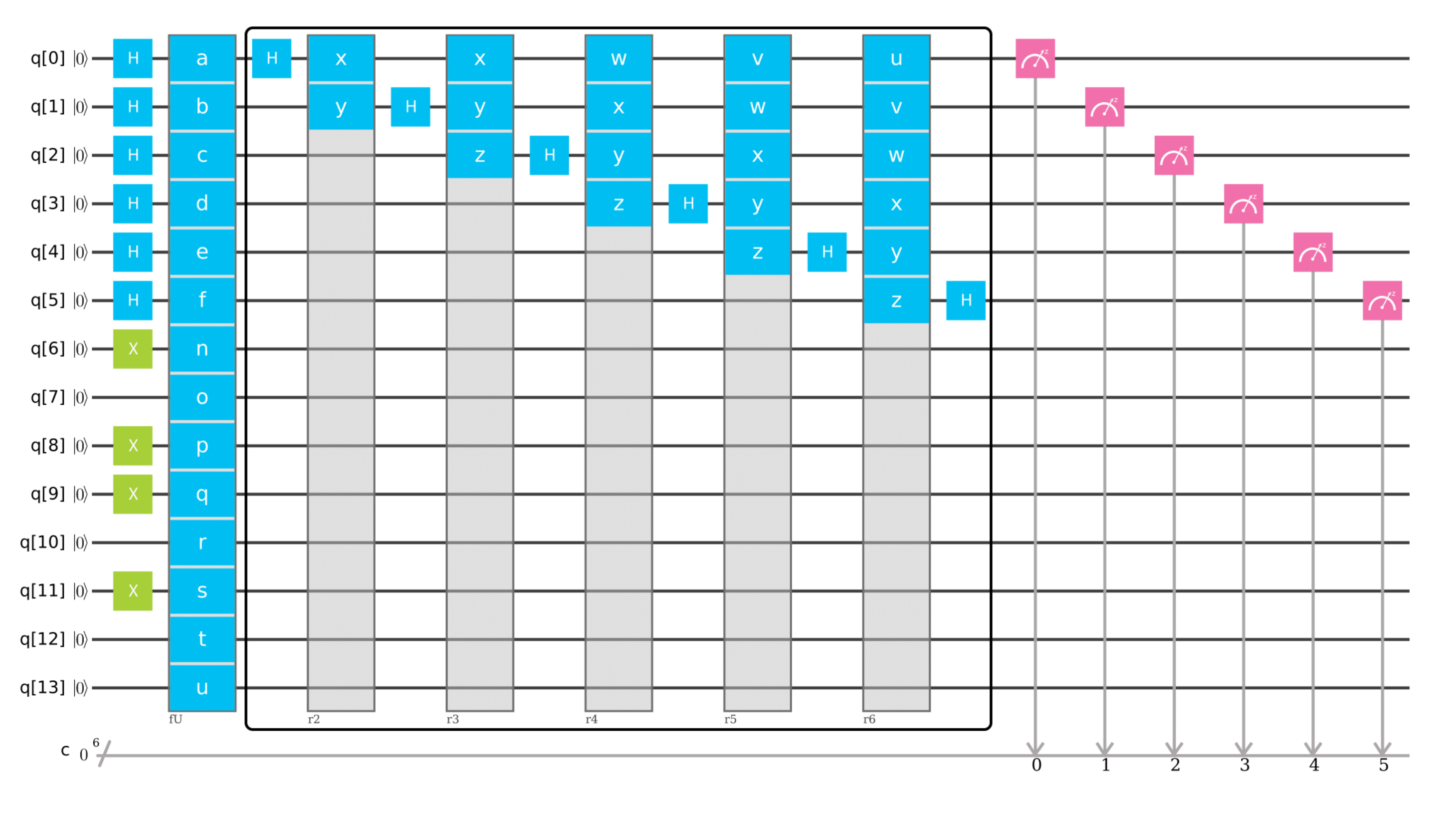}
    \caption{The figure depicts the circuit implemented in the custom topology in IBM quantum experience. The codes for the subroutines depicted in the figure are given in the qasm code for the entire circuit presented above. The circuit in the box performs inverse quantum Fourier transform for six qubits. }
    \label{tsp_Fig4}
\end{figure}


\begin{thebibliography}{10}
\expandafter\ifx\csname url\endcsname\relax
  \def\url#1{\texttt{#1}}\fi
\expandafter\ifx\csname urlprefix\endcsname\relax\def\urlprefix{URL }\fi
\providecommand{\bibinfo}[2]{#2}
\providecommand{\eprint}[2][]{\url{#2}}

\bibitem{tsp_mottnat2017}
\bibinfo{author}{Mott, B., Job, J., Vlimant, J.~R., Lidar, D. \& Spiropulu, M.}
\newblock \bibinfo{title}{Solving a Higgs optimization problem with quantum annealing for machine learning}.
\newblock \emph{\bibinfo{journal}{Nature }}
  \textbf{\bibinfo{volume}{550}}, \bibinfo{pages}{375--379}
  (\bibinfo{year}{2017}).

\bibitem{tsp_WesterlundSD1998}
\bibinfo{author}{Westerlund, T., Harjunkoski, I. \& Isaksson, J.} 
\newblock \bibinfo{title}{Solving a production optimization problem in a paper-converting mill with MILP}.
\newblock \emph{\bibinfo{journal}{Comput. Chem. Eng.}}
  \textbf{\bibinfo{volume}{22}}, \bibinfo{pages}{563--570}
  (\bibinfo{year}{1998}).

\bibitem{tsp_DebMOOP2009}
\bibinfo{author}{Deb K. \& Sinha A.}
\newblock \bibinfo{title}{Solving Bilevel Multi-Objective Optimization Problems Using Evolutionary Algorithms}.
\newblock \emph{\bibinfo{journal}{Evol. Multi-Criterion Optim.}}
  \textbf{\bibinfo{volume}{5467}}, \bibinfo{pages}{110--124}
  (\bibinfo{year}{2009}).
  
\bibitem{tsp_Wangarxiv2014}
\bibinfo{author}{Wang, Z.}
\newblock \bibinfo{title}{On Solving Convex Optimization Problems with Linear Ascending Constraints}.
\newblock \emph{\bibinfo{journal}{arXiv preprint 1212.4701v7}}
  %\textbf{\bibinfo{volume}{}}, \bibinfo{pages}{}
  (\bibinfo{year}{2014}). 

 \bibitem{tsp_Fiorettoarxiv2016}
\bibinfo{author}{Fioretto, F., Pontelli, E. \& William, Y.} 
\newblock \bibinfo{title}{Distributed Constraint Optimization Problems and Applications: A Survey}.
\newblock \emph{\bibinfo{journal}{J. Artif. Intell. Res.}}
  %\textbf{\bibinfo{volume}{}}, \bibinfo{pages}{}
  (\bibinfo{year}{2018}).

\bibitem{tsp_ZhaoJOTA2018}
\bibinfo{author}{Zhao, S. \& Zhou, J.}
\newblock \bibinfo{title}{Solutions to Constrained Optimal Control Problems with Linear Systems}.
\newblock \emph{\bibinfo{journal}{J. Optim. Theory Appl.}}
\bibinfo{pages}{1--14}
  (\bibinfo{year}{2018}).

\bibitem{tsp_ButenkoJOTA2003}
\bibinfo{author}{Butenko, S., Murphey, R., \& Pardalos, P.~M.}
\newblock \bibinfo{title}{Recent Developments in Cooperative Control and Optimization}.
\newblock \emph{\bibinfo{journal}{Springer}}
%\textbf{\bibinfo{volume}{}}, \bibinfo{pages}{}
  (\bibinfo{year}{2003}).

\bibitem{tsp_Daiarxiv2018}
\bibinfo{author}{Dai, H., Khalil, E.~B., Zhang, Y., Dilkina, B. \& Song, L.} 
\newblock \bibinfo{title}{Learning Combinatorial Optimization Algorithms over Graphs}.
\newblock \emph{\bibinfo{journal}{arXiv preprint 1704.01665v4}}
  %\textbf{\bibinfo{volume}{}}, \bibinfo{pages}{}
  (\bibinfo{year}{2018})

\bibitem{tsp_HoffmanJCAM2000}
\bibinfo{author}{Hoffman, K. L.}
\newblock \bibinfo{title}{Combinatorial optimization: Current successes and directions for the future}.
\newblock \emph{\bibinfo{journal}{J. Comput. Appl. Math.}}
  \textbf{\bibinfo{volume}{124}}, \bibinfo{pages}{341--360}
  (\bibinfo{year}{2000}).
  
\bibitem{tsp_Durrarxiv1996}
\bibinfo{author}{Durr, C., \& Hoyer, P.} 
\newblock \bibinfo{title}{A Quantum Algorithm for Finding the Minimum}.
\newblock \emph{\bibinfo{journal}{arXiv preprint quant-ph/9607014}}
 % \textbf{\bibinfo{volume}{}}, \bibinfo{pages}{}
  (\bibinfo{year}{1996}).
  
\bibitem{tsp_Kumlander2008}
\bibinfo{author}{Kumlander D.}
\newblock \bibinfo{title}{NP-Hard Graph Problems’ Algorithms Testing Guidelines: Artificial Intelligence Principles and Testing as a Service}.
\newblock \emph{\bibinfo{journal}{Innovative Tech. in Instruction Technol., E-learning, E-Assess., and Educ. Springer, Dordrecht}}
\bibinfo{pages}{112--116}
  (\bibinfo{year}{2008}).
  
 \bibitem{tsp_Herrnat2017}
\bibinfo{author}{Herr, D., Nori, F. \& Devitt, S.~J.}
\newblock \bibinfo{title}{Optimization of lattice surgery is NP-hard}.
\newblock \emph{\bibinfo{journal}{npj Quantum Inf.}}
  \textbf{\bibinfo{volume}{3}}, \bibinfo{pages}{35}
  (\bibinfo{year}{2017}). 

 \bibitem{tsp_berge1973graphs}
\bibinfo{author}{Berge, Claude \& Minieka, Edward}
\newblock \bibinfo{title}{Graphs and hypergraphs}.
\newblock \emph{\bibinfo{journal}{North-Holland publishing company Amsterdam}}
  %\textbf{\bibinfo{volume}{3}}, \bibinfo{pages}{35}
  (\bibinfo{year}{1973}). 


\bibitem{tsp_LawlerOR1966}
\bibinfo{author}{Lawler, E.~L. \& Wood, D.~E.}
\newblock \bibinfo{title}{Branch-and-Bound Methods: A Survey}.
\newblock \emph{\bibinfo{journal}{Operations Res.}}
  \textbf{\bibinfo{volume}{14}}, \bibinfo{pages}{699--719}
  (\bibinfo{year}{1966}).

\bibitem{tsp_PadbergORL1987}
\bibinfo{author}{Padberg, M. \& Rinaldi, G.}
\newblock \bibinfo{title}{Optimization of a 532-city symmetric traveling salesman problem by branch and cut}.
\newblock \emph{\bibinfo{journal}{Operations Res. Lett.}}
  \textbf{\bibinfo{volume}{6}}, \bibinfo{pages}{1--7}
  (\bibinfo{year}{1987}).
  
  \bibitem{tsp_LittleOR1963}
\bibinfo{author}{Little, J.~D.~C., Murty, K.~G., Sweeney, D.~W. \& Karel, C.}
\newblock \bibinfo{title}{An Algorithm for the Traveling Salesman Problem}.
\newblock \emph{\bibinfo{journal}{Operations Res.}}
  \textbf{\bibinfo{volume}{11}}, \bibinfo{pages}{972--989}
  (\bibinfo{year}{1963}).
  
   \bibitem{tsp_HeldOR1970}
\bibinfo{author}{Held, M. \& Karp, R.~M.} 
\newblock \bibinfo{title}{The Traveling-Salesman Problem and Minimum Spanning Trees}.
\newblock \emph{\bibinfo{journal}{Operations Res.}}
  \textbf{\bibinfo{volume}{18}}, \bibinfo{pages}{1138--1162}
  (\bibinfo{year}{1970}). 

\bibitem{tsp_LinOR1973}
\bibinfo{author}{Lin, S. \& Kernighan, W.}
\newblock \bibinfo{title}{An Effective Heuristic Algorithm for the Traveling-Salesman Problem}.
\newblock \emph{\bibinfo{journal}{Operations Res.}}
  \textbf{\bibinfo{volume}{21}}, \bibinfo{pages}{498--516}
  (\bibinfo{year}{1973}).

\bibitem{tsp_GeemSIM2001}
\bibinfo{author}{Geem, Z.~W., Kim, J.~H. \& Loganathan, G.~V.}
\newblock \bibinfo{title}{A new heuristic optimization algorithm: Harmony search}.
\newblock \emph{\bibinfo{journal}{Simulation}}
  \textbf{\bibinfo{volume}{76}}, \bibinfo{pages}{60-68}
  (\bibinfo{year}{2001}).

\bibitem{tsp_LaporteDAM1990}
\bibinfo{author}{Laporte, G., Martello, S.} 
\newblock \bibinfo{title}{The selective travelling salesman problem}.
\newblock \emph{\bibinfo{journal}{Discrete Appl. Math.}}
  \textbf{\bibinfo{volume}{26}}, \bibinfo{pages}{193-207}
  (\bibinfo{year}{1990}).

\bibitem{tsp_KargMS1964}
\bibinfo{author}{Karg, R.~L. \& Thompson, G.~L.} 
\newblock \bibinfo{title}{A Heuristic Approach to Solving Travelling Salesman Problems}.
\newblock \emph{\bibinfo{journal}{Manage. Sci.}}
  \textbf{\bibinfo{volume}{10}}, \bibinfo{pages}{225--248}
  (\bibinfo{year}{1964}).

\bibitem{tsp_BhideIEEE1993}
\bibinfo{author}{Bhide, S., John, N., \& Kabuka, M.~R.} 
\newblock \bibinfo{title}{A real -time solution for the traveling salesman problem using a boolean neural network}.
\newblock \emph{\bibinfo{journal}{IEEE Int. Conf. Neural Networks Conf. Proc.}}
\bibinfo{pages}{1096--1103)}
  (\bibinfo{year}{1993}).

\bibitem{tsp_ShaelaieASC2014}
\bibinfo{author}{Shaelaie, M.~H., Salari, M. \& Azimi, Z.~N.} 
\newblock \bibinfo{title}{The generalized covering traveling salesman problem}.
\newblock \emph{\bibinfo{journal}{Appl. Soft. Comput.}}
  \textbf{\bibinfo{volume}{24}}, \bibinfo{pages}{867-878}
  (\bibinfo{year}{2014}).

\bibitem{tsp_ShwetaIJST2017}
\bibinfo{author}{Rana, S. \& Srivastava, R.~S.}
\newblock \bibinfo{title}{Solving Travelling Salesman Problem Using Improved Genetic Algorithm}.
\newblock \emph{\bibinfo{journal}{Indian J. Sci. Technol.}}
  \textbf{\bibinfo{volume}{10}},
  (\bibinfo{year}{2017}).
  
\bibitem{tsp_Keiuarxiv2018}
\bibinfo{author}{Kieu, T.~D.}
\newblock \bibinfo{title}{The Travelling Salesman Problem and Adiabatic Quantum Computation: An Algorithm}.
\newblock \emph{\bibinfo{journal}{arXiv preprint quant-ph/1801.07859}}
  %\textbf{\bibinfo{volume}{}}, \bibinfo{pages}{}
  (\bibinfo{year}{2018}).

\bibitem{tsp_Goswamiarxiv2004}
\bibinfo{author}{Goswami, D., Karnick, H., Jain, P. \& Maji, H.~K.} 
\newblock \bibinfo{title}{Towards Efficiently Solving Quantum Traveling Salesman Problem}.
\newblock \emph{\bibinfo{journal}{arXiv preprint quant-ph/0411013}}
  %\textbf{\bibinfo{volume}{}}, \bibinfo{pages}{}
  (\bibinfo{year}{2004}).


\bibitem{tsp_BangJKPS2012}
\bibinfo{author}{Bang, J., Yoo, S., Lim, J., Ryu, J., Lee, C. \& Lee, J.} 
\newblock \bibinfo{title}{Quantum heuristic algorithm for traveling salesman problem}.
\newblock \emph{\bibinfo{journal}{J. Korean Phys. Soc.}}
  \textbf{\bibinfo{volume}{61}}, \bibinfo{pages}{1944--1949}
  (\bibinfo{year}{2012}).

\bibitem{tsp_MoylettPRA2017}
\bibinfo{author}{Moylett, D.~J., Linden, N. \& Montanaro, A.}
\newblock \bibinfo{title}{Quantum speedup of the traveling-salesman problem for bounded-degree graphs}.
\newblock \emph{\bibinfo{journal}{Phys. Rev. A}}
  \textbf{\bibinfo{volume}{95}}, \bibinfo{pages}{032323}
  (\bibinfo{year}{2017}).
  
\end{thebibliography}

\begin{thebibliography}{10}
\expandafter\ifx\csname url\endcsname\relax
  \def\url#1{\texttt{#1}}\fi
\expandafter\ifx\csname urlprefix\endcsname\relax\def\urlprefix{URL }\fi
\providecommand{\bibinfo}[2]{#2}
\providecommand{\eprint}[2][]{\url{#2}}

\bibitem{tsp_sup_IBM1}
\bibinfo{author}{IBM Quantum Experience} 
\newblock 
\emph{\bibinfo{url}{https://www.research.ibm.com/ibm-q/}}

\end{thebibliography}
\end{document}